\documentclass{aa}
\usepackage{graphicx}
\usepackage{txfonts}
\newcommand{\teff}{\mbox{${T}_{\rm eff}$}}

\newcommand{\msol}{\mbox{${\rm M}_{\odot}$}}
\newcommand{\msun}{\mbox{${\rm M}_{\odot}$}}

\newcommand{\simgt}{\lower.5ex\hbox{$\; \buildrel > \over \sim \;$}}
\newcommand{\simlt}{\lower.5ex\hbox{$\; \buildrel < \over \sim \;$}}

\begin{document}
\title{Efficiency of convection and Pre-Main Sequence Lithium depletion}
%% \subtitle{Application of new grids of ATLAS9 atmosphere models}
\author{ F.~D'Antona\inst{1} and J.~Montalb\'an\inst{2}}
\offprints{J.~Montalb\'an}
\institute{
\inst{1}INAF, Osservatorio Astronomico di Roma, I--00040 Monteporzio, Italy\\
\inst{2} Institut d'Astrophysique et de G\'eophysique, Universit\'e de Li\`ege, 4000 Li\`ege, 
Belgium}

\date{}
\titlerunning{Pre Main sequence and Lithium depletion}
\authorrunning{F. D'Antona and J. Montalb\'an}

\abstract
{We show by detailed model computation how much the Pre-Main Sequence 
(PMS) Lithium depletion depends on the treatment of over-adiabaticity, by taking 
advantage of the results of new models by Montalb\'an et al., which 
apply different treatments of convection to non-grey PMS models. 
In order to reproduce both the PMS lithium depletion (inferred from the lithium 
depletion patterns in young open clusters), and the location of PMS tracks
in the HR diagram (inferred from the study of young PMS stars),  convection both in the 
atmosphere and in a good fraction of the stellar envelope must be highly inefficient:
e.g., in the Mixing Length Theory approximation, it must have a very low $\alpha=l/H_p$. 
Unfortunately, {the radii of} these models are at variance with the solar radius, possibly indicating that 
there is some additional physical input, generally not taken into account in the stellar 
models, which affects the efficiency of convection in PMS stars, but probably not in 
the main sequence stars nor in evolved red giants.
We stress the importance of determining precisely masses and lithium abundance in PMS binaries such
as the important spectroscopic and eclipsing binary RXJ~0529.4+0041.
\keywords{stars: evolution -- stars: atmospheres -- convection -- stars: pre main sequence}}

\maketitle

\section{Introduction}

The location in the HR diagram of the Pre-Main Sequence (PMS) evolutionary 
tracks is very sensitive to physical inputs such as 
low--temperature opacity, equation of state, rotation, atmosphere model, and the 
treatment of convection. During the last years, much work has been done to 
improve the knowledge of low--temperature opacities and to include them in the 
modeling of stellar atmospheres. Among the many theoretical evolutionary 
tracks today available (new convection treatment with grey boundary conditions; 
classic convection with non-grey atmosphere models...) it is difficult to disentangle 
the effect of the different physical inputs on the results. 
In paper I (Montalb\'an et al. 2003) we made a detailed exploration of models, 
with the aim  to extricate the different roles of non-grey atmospheres and of 
convection in the track computations. This task has been approached by comparing 
a large number of models, having different assumptions for the treatment of convection both in the 
atmosphere and in the interior, and using different grids of atmospheric computations.
In particular, we have shown that a stellar 
model is fully described only when we specify not only the atmospheric model 
used as boundary condition to the interior, but also the convection parameters 
used for the atmospheric grid, and the value of optical depth at which the 
boundary conditions are taken, that is the photospheric `matching point' $\tau_{\rm ph}$. 
In fact, until now, the `expensive' model atmosphere 
computations have been  generally performed only for one specific convection model, 
e.g. a given ratio of mixing length to pressure scale heigth, in the 
Mixing Length Theory (MLT), $\alpha=\alpha_{\rm atm}$. The value of $\alpha$\ can be
changed only in the computation of the interior ($\alpha=\alpha_{\rm in}$)
e.g. to fit the solar radius in the solar model. However, if a 
large value of $\tau_{\rm ph}$\ is chosen as matching point between the 
atmosphere and the interior, the most superadiabatic part of the convection zone is all 
included in the atmosphere. Consequently, the changing of $\alpha_{\rm in}$\  
does not affect the model 
in the same way as a full change the convection parameter in the whole model,
including the atmosphere (see paper I).

Although this may seem only of academic interest, it is of subtle importance, if 
we wish to progress in one of the most intriguing problems of the stellar 
structure, namely the problem of PMS lithium depletion. In fact, the well known 
`problem of lithium in the Sun', which in the years 1965-1990 was mainly that 
solar models could not burn any substantial fraction of their initial lithium 
during the PMS, was generally taken as a good proof that additional mechanisms 
for depletion were required, acting during the long solar MS lifetime. This interpretation
is today taken as the most plausible one, confirmed by the variation of the Lithium
Depletion Patterns (LDP) in open clusters (see e.g. Chaboyer 1998). 
In fact the lithium vs. \teff\  relation for the MS stars of young open clusters 
indicates a lithium depletion by at most a factor two for the 
solar mass in young clusters, while it is compatible with the solar depletion (a 
factor $\sim$140 with respect to the solar system abundance) 
in some stars of the cluster M67, close to the solar 
age. For recent reviews see e.g. Jeffries (2000) and Pasquini (2000).

\begin{table}
\caption{Lithium remnant after the PMS depletion in 1M$_\odot$ models,
starting with $\log$ N(Li)$_{\rm in}=3.31$ }
\begin{tabular}{lc|cc|l|l}   \hline
%       \medskip
atmosphere  & $\alpha_{\rm atm}$ & interior & $\alpha_{\rm in}$  &$\tau_{\rm ph}$&log N(Li) \\
\hline
ATLAS9    MLT & 0.5  &  MLT   & 1.85  & 3      &   1.864  \\
ATLAS9  MLT   & 0.5  &  MLT   & 2.3    & 100  &   2.130  \\
ATLAS9    MLT & 0.5  &  MLT   & 6.3    & 10    &   2.564  \\
ATLAS9   FST   & 0.1  &  FST    & 0.2     & 10   &   1.537   \\
AH97   MLT      &   1.0 &  MLT  &  1.9    & 3      &  1.960  \\
AH97   MLT      &   1.0 &  MLT  &  1.9    & 100   &  2.62  \\
\hline
\end{tabular}
\label{tab1}
\end{table}

However,  a different problem emerges from the most recent computation of solar models: 
they deplete too much lithium during the PMS evolution (D'Antona and Mazzitelli 
1994, 1997, Schlattl and Weiss 1999, Piau and Turck-Chi{\` e}ze 2002) and are 
incompatible with the open clusters observations.\footnote{We mention, as a caveat, 
that these models are not strictly incompatible with the present solar lithium abundance, 
if we accept that all or most of the solar lithium was depleted at the PMS stage.} This 
problem is most severe in models using very efficient convection models, in fact 
it is more relevant for the D'Antona and Mazzitelli (1994 and 1997) models 
adopting the Full Spectrum Turbulence (FST) convection by Canuto and 
Mazzitelli (1991) and/or by Canuto et al. (1996). MLT models of the most recent 
generation, adopting updated equations of state and opacities also deplete too much lithium: 
notice however that the problem is found {mainly in tracks whose convection
is adjusted to provide the solar radius at the solar age! If one does not require
the solar fit,} it is easy to decrease the convection efficiency and to obtain a 
small PMS lithium depletion. 

A similar problem is found when comparing the location of PMS theoretical tracks with the
observed few data of PMS stars for which an independent determination of mass is available, 
either because they belong to binaries (e.g. Covino et al. 2001, Steffen et al. 2001) or by the
measure of the dynamical properties of their  protoplanetary disks
(Simon et al. 2000): the tracks most consistent with the observations are those with cooler 
atmospheres (higher mass for a given spectral type) and thus those which, generally, provide
a radius larger than R$_\odot$\ for the solar model. In particular, 
we can quote the analysis by Simon et al. (2000): their results are consistent, --within the
statistical errors-- with the set of tracks whose evolution for the solar mass
does not pass through the solar location. In addition, their observations are not
consistent with the FST tracks. In fact, for any value of its fine tuning  parameters, 
the FST convection always provides solar tracks which pass within $\sim 100$K of the solar location.

These two problems, HR diagram location of the tracks during the PMS evolution,
and lithium depletion, are necessarily correlated each other,
because, the smaller the \teff\ of the Hayashi track, the smaller will be the temperature at the base 
of the convective envelope during the possible phase of lithium burning (see, e.g. D'Antona 
et al. 2000).  
Having computed several sets of stellar models, which are fully described in paper I, 
we decided to look at the
lithium depletion with different assumptions. In Sect. 2 we show and discuss the results. 
In Sect. 3 we show that the models having low lithium depletion also reproduce better
the location of PMS stars in the HR diagram, but they do not reproduce the solar radius. 
Thus the problem is now better defined: the efficiency of convection in PMS must be highly
reduced with respect to the efficiency in MS. In particular, those tracks whose convection 
formulation {\it can not} provide a solar model with radius far different from the 
solar radius --noticeably the FST models-- can not
reproduce the location in PMS and its lithium depletion, unless the efficiency of their convective
transport is altered by another physical input acting during the PMS, but not during the
MS or post--MS giant phase.

\section{Standard models: fitting the solar radius, or not}

\begin{figure}
\resizebox{\hsize}{!}{\includegraphics{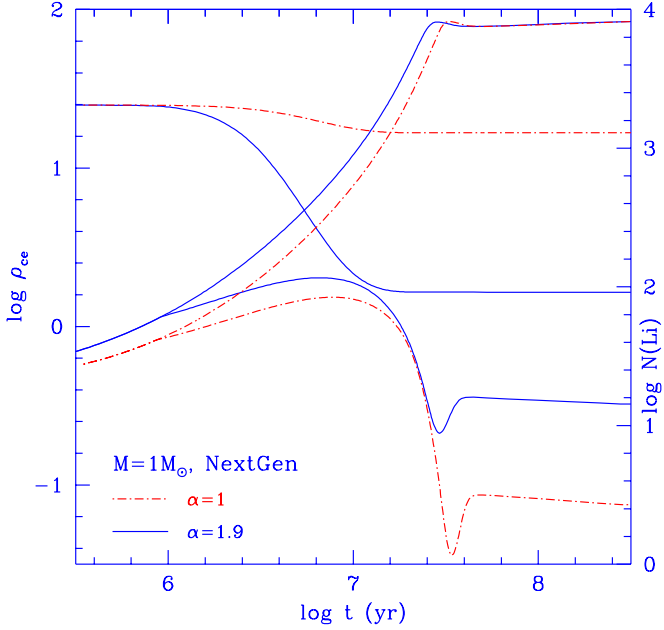}}
\caption{Along two tracks for the solar mass, differing in $\alpha_{\rm in}$, we show the 
lithium depletion (upper curves, scale on the right) and the central density and the density 
at the bottom of the convective envelope (curves which detach at $\log t\sim 6$, scale on the left).
  \label{f11}
  }
\end{figure}

Our models are computed following standard Henyey type integration, by means of the ATON2.0 code
(Ventura et al. 1998a, Montalb\'an et al. 2001) updated in paper I 
to include as surface boundary conditions (BCs) the new grids of ATLAS9 
atmospheres from Heiter et al.\ (2002a). We consider here the models having 
solar metal abundance in mass fraction Z=0.02 and helium abundance Y=0.28.
The initial lithium abundance was taken as
log N(Li)=3.31$\pm 0.04$, the solar system abundance given by Anders and Grevesse 1989. 
The PMS lithium burning starts when the temperature reaches $\sim 2 \times 10^6$K during the
PMS contraction. The important inputs are the temperature and density at the bottom of the 
convective region. Actually lithium burning generally begins approximately at the bottom of the
Hayashi track, when a radiative core develops and grows during the evolution towards the main 
sequence. Figure \ref{f11} shows the the central density, the density at the bottom of the
convective region, and the surface lithium abundance along two tracks of 1$M_\odot$, whose atmosphere
is interpolated in the grid of models by Allard and Haushildt (1997). Varying the convection 
treatment in the interior from $\alpha_{\rm in}=l/H_p=1$\ to 1.9, increases the temperature at the
bottom of the envelope by  $\sim 3$\%. As we see, the density is instead increased by $\sim$50\%.
The difference in the structure is sufficient to change the lithium depletion by a 
factor $\sim$10. 
As shown in Fig. \ref{f11}, the lithium burning phase is completed when the
age is still as low as 10$^7$yr. These models do not include mechanisms for depletion other than 
plain convective mixing. These latter, however, act on a much longer 
timescale during the MS phase, so that the remaining
lithium abundance should be preserved at least until the age of the Pleiades (see,
e.g. Pinsonneault et al. 1990). Notice that
lithium burning begins approximately when the radiative core develops (in Fig. \ref{f11}, 
where the central density
curve detaches from the density at the bottom of convection). This occurs at the knee in the
HR diagram, where the track shifts from its convective behavior (the Hayashi track) to the
so called Henyey --radiative-- line of approach to the MS (e.g. Stahler 1988). Thus the lithium
burning begins at a \teff\ around the minimum luminosity of the track, which we tentatively  
use as a possible interesting parameter. This $T_{\rm eff-lim}$\ gives us 
an indication of the PMS track location in the HR diagram (being smaller for
a smaller efficiency of convection) and we expect that it might be correlated with the
PMS lithium burning.

Table \ref{tab1} shows the remaining lithium abundance in the solar tracks after the
 PMS depletion, in several models from paper I, all of which 
 fulfilling the requirement that the solar radius is achieved at the solar age. 
The models are either computed by using the new Vienna ATLAS9 (Kurucz 1993, 1995, 1998) 
model atmospheres
(Heiter et al. 2002) or the NextGen models by Allard and Hauschildt (1997, AH97).
Among the ATLAS9 models, some are computed by the Full Spectrum Turbulence (FST)
convection description by Canuto et al. (1996) both in the atmosphere and in the interior.
We see that  depletion is very severe in the FST model. For the 
MLT models computed with the Vienna ATLAS9 model atmospheres (Heiter et al. 
2002) which have $\alpha_{\rm atm}=0.5$, the solar fit is achieved by assuming a larger 
$\alpha_{\rm in}$ in the interior computation.  The larger the $\tau_{\rm ph}$ at which
the atmosphere ends, the larger is the $\alpha_{\rm in}$\ required for the solar fit,
and the smaller the lithium depletion. In fact, the larger is $\tau_{\rm ph}$, the larger is
the fraction of the over-adiabatic region integrated using a low efficiency of convection,
so that the \teff\ of the model is significantly reduced, as well as the temperature
 at the bottom of the convective envelope at the epoch of lithium burning.
Nevertheless, the {\it minimum} depletion achieved in these
models is $\sim 0.6$~dex, still too much to be compatible with the open cluster LDPs 
(see later). Thus we confirm that
 these models using non--grey boundary conditions have
a strong lithium depletion, as it is obtained in the grey MLT models (e.g. Schlattl and Weiss
1999, Piau and Turck-Chi{\` e}ze 2002) as well as in the grey 
FST models (D'Antona and Mazzitelli 1994, 1997).

Table 2 shows the lithium depletion for masses from 1.1~\msun\ to 0.7~\msun\ in 
two sets of MLT models computed using Allard and Hauschildt (1997) NextGen model 
atmospheres as boundary condition.  All these models consider
 $\alpha_{\rm atm}=1$, the value fixed in the grid of model atmospheres,  
 but one set is computed assuming  $\alpha_{\rm in}=1.9$
in the interior, so that the solar track fits the solar radius, 
and in the other set we adopt the same value 
than in the atmosphere, $\alpha_{\rm in}=1$, even if the corresponding evolutionary track
does not fit the Sun. These latter models, whichever is the matching point  $\tau_{\rm ph}$,
have very similar lithium depletion\footnote{If the interior and atmospheric physics 
were exactly the same, we should obtain the same model structure, at least for 
the $\tau_{\rm ph}=10$ and 100 models (at $\tau_{\rm ph}=3$ the diffusion approximation 
is not yet strictly valid and this may give some differences). Actually,
also numerical differences can be responsible for some variation in the
results: in fact the atmospheric structure in the AH97 model atmospheres includes
only very few mesh points between $\tau =10$ and 100, while the interior integration is much more
detailed.}.

\begin{table}
\caption{Lithium depletion in NextGen based models with $\alpha=1$\ and $\alpha=1.9$ in the interior.}
\begin{tabular}{cc|c|l|l}   \hline
%       \medskip
M/M$_\odot$ &$\alpha_{\rm in}$ & $\tau_{\rm ph}$ & $\log T_{\rm eff-lim}$  &log N(Li) \\
\hline
 1.1   &  1.9    &       3    &   3.663  &   2.93         \\
 1.1   &  1.9    &       10   &   3.653  &   2.93         \\
 1.1   &  1.9    &       100  &   3.650  &   2.93         \\
 1.1   &  1      &       3    &   3.619 &     3.27       \\
 1.1   &  1      &       10   &   3.615  &    3.27        \\ 
 1.1   &  1      &       100  &   3.613  &    3.27        \\ 
\hline                                                          
 1     &  1.9    &      3     &   3.649  &   1.96         \\ 
 1     &  1.9    &      10    &   3.637  &   2.52         \\ 
 1     &  1.9    &      100   &   3.633  &   2.62         \\ 
 1     &  1.0    &      3     &   3.604  &   3.111        \\ 
   1   &  1.0    &        10  &     3.609&     3.249      \\ 
   1   &  1.0    &        100 &     3.607&     3.236      \\ 
\hline
   0.9 &  1.9    &        3   &     3.631&     0.845      \\ 
   0.9 &  1.9    &        10  &     3.619&     1.920      \\ 
   0.9 &  1.9    &        100 &     3.615&     2.118      \\ 
   0.9 &  1.0    &        3   &     3.597&     2.880      \\ 
   0.9 &  1.0    &        10  &     3.593&     2.980      \\ 
   0.9 &  1.0    &        100 &     3.595&     2.930      \\ 
 \hline                                                          
   0.8 &  1.9    &        3   &     3.604&     0.85       \\ 
   0.8 &  1.9    &        10  &     3.609&     1.93       \\ 
   0.8 &  1.9    &        100 &     3.607&     2.12       \\ 
   0.8 &  1.0    &        3   &     3.582&     2.345      \\ 
 \hline
   0.7 &  1.9    &        3   &     3.604&     -3.36      \\ 
   0.7 & 1.9    &        100 &     3.609&     -1.49       \\ 
 \hline

\end{tabular}                                     
\label{tab2}  
\end{table}

\begin{figure}
\resizebox{\hsize}{!}{\includegraphics{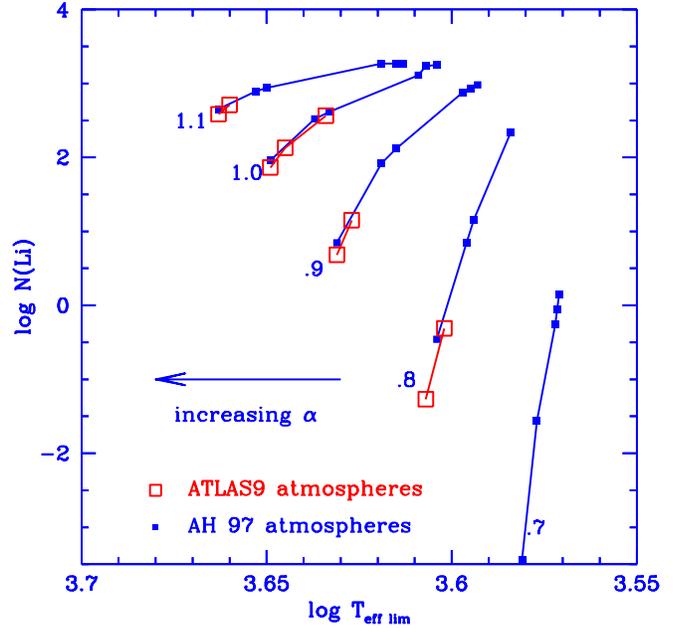}}
\caption{Lithium versus $T_{\rm eff-lim}$ relation for tracks from 1.1 to 0.7\msol, 
as labeled in the figure.
The full dots are the models computed with AH97 model atmospheres, while the 
open squares employ the ATLAS9 MLT Vienna model atmospheres by Heiter et al.\ 
(2002). The 0.8\msol models of $\alpha_{\rm atm}=\alpha_{\rm in}=1$ coincide at 
the top of the sequence of 0.8\msol models, while they differ slightly 
for the other masses, due to possible differences in the input physics of the 
model atmosphere and interior. }
\label{f1}
\end{figure}

In Table \ref{tab2} we also list the parameter  
$T_{\rm eff-lim}$, and Fig. \ref{f1} shows the lithium depletion versus
$T_{\rm eff-lim}$ : for large enough masses (M$\simgt$1.1~\msun)
the depletion is small for any convection efficiency, but the \teff\ location of the track
increases with the convection efficiency ($T_{\rm eff-lim}$ increases). 
For the lowest masses (0.7~\msun) the \teff\
location does not vary significantly, 
but the lithium depletion varies by several orders of magnitude from
 the models with low to those with high convection efficiency. 
 The masses 0.8--1\msun\ show a wider range of variability, and a very significant
 degree of correlation between $T_{\rm eff-lim}$\ and the lithium depletion.
 
\begin{figure}
\resizebox{\hsize}{!}{\includegraphics{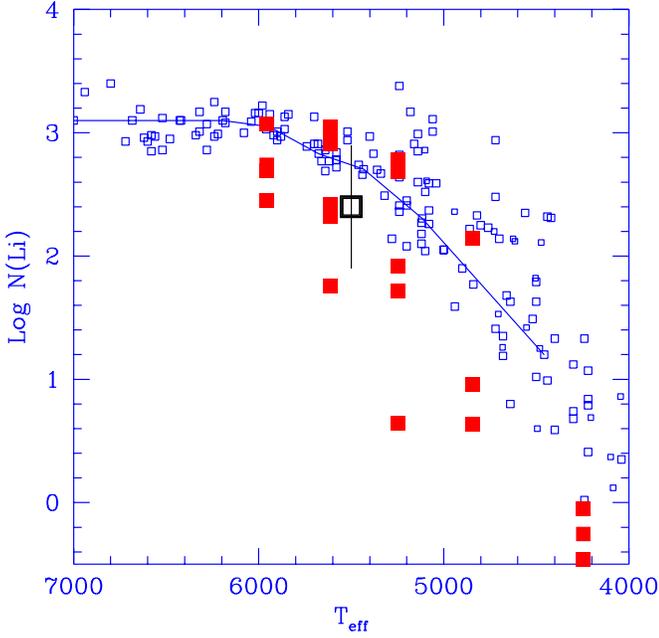}}
\caption{The Pleiades data by Soderblom et al. (1993) and Garcia Lopez et al. (1994)
(open squares) are compared with the
depletion predicted by the models in Table 2 (full big squares). The models are placed
at the \teff\ they would take in an empirical MS, at the Pleiades age. Only the upper squares,
corresponding to the models with  $\alpha_{\rm atm}=\alpha_{\rm in}=1$, are compatible 
with the data. The full line shows the depletion from the  models by Ventura et al. (1998b)
computed including the thermal effect of a magnetic field on the convective temperature 
gradients. 
 The large open square with the error bar represents the lithium abundance of the secondary component of
 RXJ~0529.4+0041  (log N(Li)=2.4$\pm$0.5, Covino et al. 2001). The \teff\ at which the point is located
 (5500K) is assumed to be the main sequence \teff\ of a star of mass 0.925\msol.
 }
 
 \label{f2}
\end{figure}

Fig.~\ref{f1} also shows the points corresponding to the ATLAS9 MLT  models fitting the Sun: these
have very similar lithium depletion, for the  solar track, to that of the AH97 
models also fitting the Sun. 
 For the other masses, the results differ,  but they follow the same pattern
 along the proper relation lithium -- $T_{\rm eff-lim}$. We did not consider ATLAS9 models 
 not fitting the Sun, but, if we had assumed such a low efficiency of  convection also in 
 the interior, we would have obtained smaller lithium depletions, as shown by the AH97 models
 with $\alpha =1$\ both in the atmosphere and in the interior. 
 We immediately recognize that the lithium depletion in young open clusters requires
 the {\it minimum}  efficiency of convection among those we have considered:
 only the models with AH97 model atmospheres
 and $\alpha_{\rm atm}=\alpha_{\rm in}=1$\ can fit the open cluster lithium data, {\it although the
 tracks do not fit the solar location}. 
 This is shown in Fig. \ref{f2} where we compare the lithium
depletion predicted by the models in Table 2 with
 the Pleiades data by Soderblom et al. (1993) and Garcia Lopez et al. (1994).
 We have normalized the lithium abundances in the models to an initial abundance log N(Li)=3.1.
 We see that only upper squares, corresponding to the models with  $\alpha_{\rm atm}=\alpha_{\rm in}=1$, 
are compatible  with the data. In addition, all the models of 0.7\msun\ 
have depletion larger than allowed by the data, suggesting that some additional
phenomenon is taking place at lower \teff. In the figure we also plot (full line)  
the depletion from the  models by Ventura et al. (1998b) computed including the thermal 
effect of a dynamo induced magnetic field on the convective temperature 
gradients. In this latter case, the magnetic field acts to reduce the convective 
efficiency, allowing the fit with the cluster LDP.
  We conclude that PMS  depletion is compatible {\it only}
 with models in which {\it PMS convection is much less efficient than MS convection}.

\begin{figure}
\resizebox{\hsize}{!}{\includegraphics{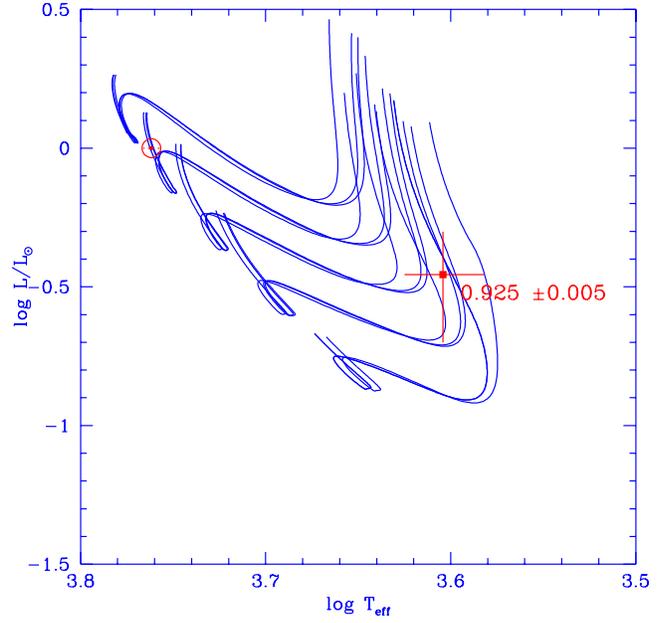}}
\resizebox{\hsize}{!}{\includegraphics{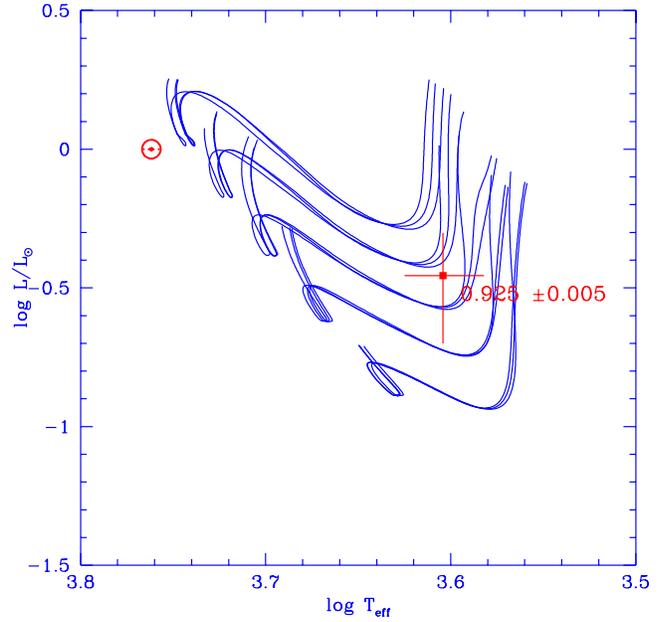}}
\caption{The top figure shows the tracks computed with AH97 atmospheres and $\alpha _{\rm in}=1.9$,
 while the bottom figure shows the tracks with $\alpha _{\rm in}=1.0$. For each mass (1.1, 1.0, 0.9,
0.8 and 0.7\msun, from top to bottom) the three curves represent the tracks obtained for different
$\tau _{\rm ph}$. The solar location is shown: it is compatible with the solar
model having $\alpha _{\rm in}=1.9$,
but $\sim$400K {\it hotter} than the solar model with $\alpha _{\rm in}=1.0$. 
The location of the secondary component of the binary RXJ~0529.4+0041 is also shown.}
\label{f3}
\end{figure}

\section{Constraints on the PMS location of tracks}

We have mentioned that the (scarce) data on PMS 
masses seem to indicate that the tracks most compatible, in terms of \teff\ and luminosity,
with the observations 
are those in which convection is less efficient (Covino et al. 2001, Steffen et al. 2001,
Simon et al. 2000). A valid addition to the analysis can be the consideration of the Lithium content. 
An example of how the analysis could be done is shown by means of
the only relevant observation available, concerning the 
secondary of the eclipsing double-lined spectroscopic binary, containing both 
components in PMS,  RXJ~0529.4+0041 (Covino et al.\ 2000). This system provides powerful
constraints: it is a very important
indicator of the  fundamental stellar parameters, as masses and 
radii of the components are simultaneously derived. Fig. 3 shows the location of the secondary of 
RXJ~0529.4+0041 in the HR diagram, compared with  the tracks having convection 
parameters fitting the Sun (upper panel) or not fitting the Sun (lower panel). 
The mass of this star is given as 0.925$\pm0.005$\msol\ in Covino et al. (2001). We 
see that both sets of tracks are compatible with this mass within the errors. 
The tracks of 1 and 0.9\msun\
in the sets not fitting the Sun very nicely bracket the star location and seem most 
adequate. Notice that in the following we 
rely on the assumption that both the Covino et al. 2001 data and the results concerning
the HR diagram location of PMS objects can be taken at face value. Actually, the errors on
the \teff\ determination might result to be larger than given by the authors, and for this reason we
judge that a more quantitative analysis is still premature.
The lithium abundance determination by Covino et al. (2000), $\log$ N(Li)=2.4$\pm 0.5$ is shown in 
Fig. \ref{f2}. As the star may still be at a phase in which PMS depletion is 
going on, this determination is an upper limit to the future location of the star in the
[lithium {\it vs.} MS--\teff] plane. According to the authors, actually the abundance determination 
should be regarded as a lower limit, as it was done by using the Pavlenko and Magazz\`u (1996)
model atmospheres for a gravity log g=4.5, more appropriate for main sequence gravities. Reduction
of the gravity would thus provide a larger abundance.
In spite of its large error, this value, considered as a lower limit, is therefore
well compatible with the quite low depletion of the $\alpha_{\rm in}=1$\ models
and with the average lithium depletion expected from the observations in young clusters.
{\it Further studies of this system, and in particular a better determination of its
lithium abundance, are important to help us in discriminating how inefficient convection
must be to be compatible with these data.}

\section{Conclusions}
Although the parametrizations used for convection are highly unsatisfactory, we 
have shown  that there are two different independent indications which imply that 
convection in the PMS phase must be highly inefficient:
\begin{enumerate}
\item the location in the HR diagram of the few objects for which the mass (or an upper
value of the mass) is known, according to several comparisons appearing in the most recent 
astronomical literature (although we have still to remind that 
further work is necessary to understand 
carefully the \teff\ scale of PMS objects).
\item the PMS depletion of lithium, according to the well known observations of
LDPs in young open clusters, plus the determination of the lithium abundance in 
the secondary component of the system  RXJ~0529.4+0041, whose mass is known with 
high precision.
\end{enumerate}
We have also shown that the PMS location of the tracks in the HR diagram 
and the corresponding lithium
depletion during the PMS stage are correlated, for those masses in which 
lithium depletion is significant but not complete. Any effort to 
increase our knowledge of the PMS masses and/or of the lithium content in young 
stars will add further weight to our main conclusion, which is that {\it we can 
not parametrize the efficiency of convection in the same way in the MS and in 
the PMS}. A potential implication of this result, preliminary explored in some previous work
(Ventura et al. 1998b, D'Antona et al. 2000) is that there might be an additional 
parameter playing a role in this game, connected with the fast rotation of the 
PMS stars. 

\begin{acknowledgements}
We acknowledge support from the Italian Space Agency ASI under the contract 
ASI I/R/037/01. J.M. also acknowledge the support of Osservatorio Astronomico di Roma.
\end{acknowledgements}

{}


\begin{thebibliography}{}
\bibitem{}Alexander D.R., Ferguson J.W., 1994, ApJ 437, 879
\bibitem{}Allard F., Hauschildt P., 1997 (AH97) NextGen
\bibitem{}Anders E., Grevesse N., 1989, Geochim. Cosmochim. Acta 53, 197
%\bibitem{}Baraffe I., Chabrier G., Allard F., Hauschildt P., 2002, A\&A 382, 563
%\bibitem{}Baraffe I., Chabrier G.,1998, A\&A 327, 1054 
%\bibitem{}Bernkopf J., 1998, A\&A 332, 127
\bibitem{}B\"ohm-Vitense  E. 1958 Z. Astrophys. 46, 108
%\bibitem{}Canuto V.M., Mazzitelli I., 1991, ApJ 370, 295
%\bibitem{}Canuto V.M., Mazzitelli I., 1992, ApJ 389, 724
\bibitem{}Canuto V.M., Goldman I., Mazzitelli I., 1996, ApJ 473, 550
\bibitem{}Chaboyer B., 1998, In: Deubner F.-L, Christensen-Dalsgaard J.,
Kurtz D. (eds.) `New eyes to see inside the sun and stars'. IAU Symp. 185. p. 25
\bibitem{}Covino E., Melo C., Alcal\'a J.M., Torres G., Fern\'andez M., Frasca A., 
and Paladino R. 2001, A\&A  375, 130
\bibitem[D'Antona(2000)]{2000ESASP.445..161D} D'Antona, F.\ 2000, Star 
formation from the small to the large scale.~ESLAB symposium (33 : 1999 : 
Noordwijk, The Netherlands).~Edited by F.~Favata, A.~Kaas, and A.~Wilson.~ 
Proceedings of the 33rd ESLAB symposium on star formation from the small to 
the large scale, ESTEC, Noordwijk, The Netherlands, 2-5 November 1999 
Noordwijk, The Netherlands: European Space Agency (ESA), 2000.~ESA SP 445., 
p.161, 445, 161 
\bibitem{}D'Antona F., Mazzitelli I., 1994, ApJS 90, 467 (DM94)
\bibitem{}D'Antona F., Mazzitelli I., 1997, in ``Cool Stars 
in Clusters and Associations",
eds. G. Micela and R. Pallavicini, Mem. S.A.It. 68,807
\bibitem[D'Antona, Ventura, \& Mazzitelli(2000)]{2000ApJ...543L..77D} 
D'Antona, F., Ventura, P., \& Mazzitelli, I.\ 2000, ApJL, 543, L77 
\bibitem{}D'Antona F., Montalb\'an J., Kupka F., Heiter U., 2002, ApJ 564, L93
%\bibitem{}Fuhrmann K., Axer M., Gehren T., 1993, A\&A 271, 451
\bibitem[Garcia Lopez, Rebolo, \& Martin(1994)]{1994A&A...282..518G} Garcia 
Lopez, R.~J., Rebolo, R., \& Martin, E.~L.\ 1994, A\&A, 282, 518 
\bibitem{}Heiter U., Kupka F., van't Veer-Menneret C., Barban C.,
	  Goupil M.J., Garrido, R., 2002a, A\&A, 392, 619
  \bibitem[Jeffries(2000)]{2000scac.conf..245J} Jeffries, R.~D.\ 2000, ASP 
Conf.~Ser.~198: Stellar Clusters and Associations: Convection, Rotation, 
and Dynamos, 24
\bibitem{}Kurucz, R.L. 1998, http://cfaku5.hardvard.edu/
\bibitem{}Kurucz, R.L. 1995, CD-ROM No 13, revised
\bibitem{}Kurucz, R.L. 1993, ATLAS9 Stellar Atmosphere Programs and 2 km/s grid
   (Kurucz CD-ROM No 13)
\bibitem[Mazzitelli \& Moretti(1980)]{1980ApJ...235..955M} Mazzitelli, 
I.~\& Moretti, M.\ 1980, ApJ, 235, 955 
%\bibitem{}Montalb\'an, J., D'Antona F., Mazzitelli I., 2000, A\&A 360, 935
\bibitem{}Montalb\'an, J., Kupka F., D'Antona F., Schmidt W., 2001, A\&A 370, 982
\bibitem{}Montalb\'an, J., D'Antona F., Kupka, F., Heiter, U. 2003, A\&A (submitted)
(paper I)
\bibitem[Pavlenko \& Magazzu(1996)]{1996A&A...311..961P} Pavlenko, Y.~V.~\& 
Magazzu, A.\ 1996, A\&A, 311, 961 
\bibitem[Pasquini(2000)]{2000IAUS..198..269P} Pasquini, L.\ 2000, IAU 
Symposium, 198, 269 
\bibitem[Piau \& Turck-Chi{\` e}ze(2002)]{2002ApJ...566..419P} Piau, L.~\& 
Turck-Chi{\` e}ze, S.\ 2002, , 566, 419 
\bibitem[Pinsonneault, Kawaler, \& Demarque(1990)]{1990ApJS...74..501P} 
Pinsonneault, M.~H., Kawaler, S.~D., \& Demarque, P.\ 1990, ApJS, 74, 501 
\bibitem[Schlattl \& Weiss(1999)]{1999A&A...347..272S} Schlattl, H.~\& 
Weiss, A.\ 1999, A\&A, 347, 272 
\bibitem[Simon, Dutrey, \& Guilloteau(2000)]{2000ApJ...545.1034S} Simon, 
M., Dutrey, A., \& Guilloteau, S.\ 2000, ApJ, 545, 1034 
\bibitem[Soderblom et al.(1993)]{1993AJ....106.1059S} Soderblom, D.~R., 
Jones, B.~F., Balachandran, S., Stauffer, J.~R., Duncan, D.~K., Fedele, 
S.~B., \& Hudon, J.~D.\ 1993, AJ, 106, 1059 
\bibitem[Stahler(1988)]{1988PASP..100.1474S} Stahler, S.~W.\ 1988, PASP, 
100, 1474 
\bibitem[Steffen et al.(2001)]{2001AJ....122..997S} Steffen, A.~T.~et al.\ 
2001, AJ, 122, 997 
\bibitem[Ventura, Zeppieri, Mazzitelli, \& 
D'Antona(1998)]{1998A&A...334..953V} Ventura, P., Zeppieri, A., Mazzitelli, 
I., \& D'Antona, F.\ 1998a, A\&A, 334, 953 
\bibitem[Ventura, Zeppieri, Mazzitelli, \& 
D'Antona(1998)]{1998A&A...331.1011V} Ventura, P., Zeppieri, A., Mazzitelli, 
I., \& D'Antona, F.\ 1998b, A\&A, 331, 1011 
\end{thebibliography}
\end{document}